\def\Bbb#1{{\bf #1}}

\def\fnote#1{\footnote}
\def\blacksquare{\hbox{\vrule width 4pt height 4pt depth 0pt}}

\def\cwleftpar#1#2{\leftskip #1 \rightskip #2 plus 1fill}
\def\cwrightpar#1#2{\leftskip #1 plus 1fill \rightskip #2}
\def\cwcenterpar#1#2{\leftskip #1 plus 1fill \rightskip #2 plus 1fill}
\def\cwfullpar#1#2{\leftskip#1\rightskip#2}

\def\cwoutdent#1#2{\llap{\hbox to #1{#2 \hss}}\ignorespaces}
\def\cwparbegin#1#2#3#4#5{
	\ifcase #1 \cwleftpar{#2}{#3}
	\or \cwrightpar{#2}{#3}
	\or \cwcenterpar{#2}{#3}
	\else \cwfullpar{#2}{#3}\fi
	\ifcase #4 \baselineskip = 1.5\baselineskip
	\or \baselineskip = 2\baselineskip
	\or \baselineskip = 3\baselineskip
	\else \baselineskip = 1\baselineskip\fi
	\ifdim #5 > 0in \else \noindent \fi
	\noindent\ignorespaces}
\documentclass{article}
\begin{document}
\advance \vsize by -2\baselineskip
\def\makeheadline{
\ifnum\pageno = 1{\vskip \baselineskip \vskip \baselineskip }\fi}
\advance \vsize by -1\baselineskip
\def\makefootline{
\ifnum\pageno = 1{\vskip \baselineskip \vskip \baselineskip }\else{\vskip \baselineskip \noindent \folio                                 \par
}\fi}

 \vspace*{3ex}

\noindent {\Huge Centre of Mass in Spaces with\\[1ex] Torsion
		 Free  Flat  Linear\\[1ex] Connection}

 \vspace*{3ex}

\noindent Bozhidar Zakhariev Iliev
\fnote{0}{\noindent $^{\hbox{}}$Permanent address:
Laboratory of Mathematical Modeling in Physics,
Institute for Nuclear Research and \mbox{Nuclear} Energy,
Bulgarian Academy of Sciences,
Boul.\ Tzarigradsko chauss\'ee~72, 1784 Sofia, Bulgaria\\
\indent E-mail address: bozho@inrne.bas.bg\\
\indent URL: http://theo.inrne.bas.bg/$\sim$bozho/}

 \vspace*{2ex}

{\bf \noindent Published: Communication JINR, E5-95-159, Dubna, 1995}\\[1ex]
\hphantom{\bf Published: }
http://www.arXiv.org e-Print archive No.~gr-qc/0404002\\[3ex]

2003 PACS numbers: 04.90.+e, 45.90.+t\\[2ex]

\noindent
{\small
The \LaTeXe\ source file of this paper was produced by converting a
ChiWriter 3.16 source file into
ChiWriter 4.0 file and then converting the latter file into a
\LaTeX\ 2.09 source file, which was manually edited for correcting numerous
errors and for improving the appearance of the text.  As a result of this
procedure, some errors in the text may exist.
}\\[2ex]

	\begin{abstract}
The concept "centre of mass" is analyzed in spaces with torsion free flat
linear connection. It is shown that under sufficiently general conditions it
is almost uniquely defined, the corresponding arbitrariness in its definition
being explicitly described.
	\end{abstract}\vspace{3ex}

 {\bf 1. INTRODUCTION}
\nopagebreak

\medskip
In a series of works, which began with [2,3] and is partially summarize in [4], W. G. Dixon developed some methods of dynamics of extended bodies in general relativity. He made an essential usage of the theory of bitensors, first of all for defining some dynamical quantities in curved spaces and their further treatment.

The bitensors theory, originally considered by H.S. Ruse [8] and J. L. Synge
[9,10], is deeper investigated in [11], where some its physical applications
can be found. This theory is also widely used in [1]. It should be emphasized
that in all mentioned physical applications of bitensors mainly are used ones
obtained by differentiation of the world function, primary introduced by J.
L. Synge [10].

 The present work, which was inspired by the above references and some purely
mathematical considerations, begins an investigation of dynamics in,
generally curved, space-times endowed with a structure called a ("parallel")
"transport" (long paths), which, when it is linear and acts in the tensor
bundles over a given manifold, is equivalent to (a system of) bitensors with
a suitable properties $(cf. [13])$. In particular, here we shall use "flat
linear transports" in tensor bundles over the space-time [12] which, as it is
proved in [12], are simply parallel transports generated by flat linear
connections in these bundles. On this basis our work is aimed to analyze the
concept "centre of mass" of a physical system described with its
energy-momentum tensor and to propose an adequate definition of that concept.
The generalizations of the presented here results to the case of more general
space-times with arbitrary curvature and torsion will be published elsewhere.

In section 2, by means of flat linear transports over a given space-time, we introduce the needed for us dynamical quantities and present a part of their properties. These quantities are similar to the classical ones and coincide with them in the corresponding special case.

In section 3 we define the mass centre of a discrete physical system and consider its connection with some dynamical quantities depending on the energy-momentum tensor of this system.

Section 4 contains analysis of the mass centre of a physical system described by its energy-momentum tensor. As a ground are taken two conditions: (a) in the discrete case one must obtain the results of section 3 and (b) some linear conditions (see (4.3)) are assumed
to hold. It turns out that they define the mass centre up to an arbitrary 1-form (covector) which, when, as usual, the space-time is endowed with a metric, is naturally to be assumed to be the covector corresponding with respect to the metric to the energy-momentum vector of the system.

 Section 5 are presented certain concluding remarks.

\medskip
 {\bf 2. SOME MECHANICAL QUANTITIES DEFINED BY MEANS\\
		OF FLAT LINEAR TRANSPORTS}
\nopagebreak

\medskip
In this section certain necessary for our investigation quantities are defined and some their properties are established.

Let $M$ be a differentiable manifold [7] endowed with a flat linear transport $L [12]$, which can equivalently be thought as a parallel transport generated by a flat linear connection $\nabla ^{L}$on $M [12]$. Physically $M$ will be interpreted as a space-time of dimension $n:=\dim(M)=4$ and its properties will be specified, when needed, below.

The Latin and Greek indices are referring to $M$ and will run, respectively, from 0 to $n-1=3$ and from 1 to $n-1=3.$The usual summation rule over repeated on different levels Latin (resp. Greek) indices from 1 to $n ($resp. $n-1)$ will be assumed.

The (flat linear) transport from $x$ to $y, x,y\in M$ will be denoted by $L_{x  \to y}$and $H^{i}_{.j}(y,x)$ will mean the components of the matrix representing it (in some local coordinates), which are components of a bivector (vector at $y$ and covector at $x) [12]$. For details concerning flat linear transports the reader is referred to [12].

{\bf Definition 2.1.} Let the $C^{1}$path $\gamma :J  \to M, J$ being an
${\Bbb R}$ interval, joints the points $x,y\in M$, i.e. $\gamma (s)=x$ and
$\gamma (t)=y$ for some $s,t\in $J. The displacement vector of $y$ with
respect to $x$ (as it is defined by the transport $L)$ is the vector
\[
 h(x,y):=
\int^{t}_{s}(L_{\gamma (r)  \to \gamma (t)}\dot\gamma(r))dr,\qquad (2.1)
\]
where is the tangent to $\gamma $ vector field.

In the general case $h(x,y)$ depends on $\gamma $. We didn't denote this because hereafter in this work we shall be interested only in the case when $h(x,y)$ doesn't depend on $\gamma $. This assumption puts a restriction on the used transport $L$ which is expressed by

{\bf Proposition 2.1.} If the points $x$ and $y$ belong to some coordinate neighborhood, then the displacement vector (2.1) doesn't depend on the path $\gamma $ if and only if the torsion of the flat linear connection,
for which $L$ is a parallel transport, is zero.

  {\bf Proof.} In a coordinate basis the components of (2.1) are
\[
 h^{i}(x,y)
=\int_{s}^{t} H^{i}_{.j}(x,\gamma (r))\dot\gamma^{j}(r)dr
=\int_{x}^{y} H^{i}_{.j}(x,z)dz^{j},\qquad (2.2)
\]
 where we have made the substitution
$z^{j}=\gamma ^{j}(r)$ and the last integral is along $\gamma $. As it is
well known [6], this last integral is locally independent from $\gamma $
iff the integrand in it is a full  differential  (with respect to $z)$, i.e.
iff locally  it  is  a  closed  1-form  which  is expressed by $eq.
(4.3^\prime )$ of [12]. By its turn this  equation,  due  to proposition 4.3
and the remark after proposition 4.2   from  [12]  is satisfied iff the
mentioned torsion vanishes.\blacksquare

{\bf Remark.} If there does not exists a coordinate neighborhood containing $x$ and $y$, then the vector (2.1) depends on the path $\gamma  ($see below the remark after (2.6)). That is why further is supposed the points defining some displacement vector to belong to some coordinate neighborhood.

{ \bf Proposition 2.2.} If $x,y$ and $z$ belong to one and the same
coordinate neighborhood, it is valid the implication
\[
 h(x,y)=h(x,z)  \Leftrightarrow  y=z \qquad (2.3)
\]
which is equivalent to
\[
h(x,y)=0 \Leftrightarrow  y=x.  \qquad (2.3^\prime )
\]

 {\bf Proof.} By propositions  4.1 and 4.2 from [12] there exists
holonomic coordinates $\{x^{i^\prime }\}$ in the neighborhood containing
$x, y$ and  $z$ such that $H^{i^\prime }_{..j^\prime }(x,y)=\delta ^{i^\prime
}_{j^\prime }$. So, in it, we have
\[
h^{i^\prime }(x,y)
=\int^{y}_{x } H^{i^\prime }_{..j^\prime }(x,u)du^{j^\prime}
=\int^{y}_{ x }du^{i^\prime }=y^{i^\prime }-x^{i^\prime },\qquad (2.4)
\]
 from
where immediately follow (2.3) and $(2.3^\prime ).\blacksquare $

{\bf Remark.} From (2.4) we infer that in the considered case $h(x,y)$ is a straightforward generalization of the Euclidean (difference of two) radius-vector(s).

From this proposition, evidently, can be concluded that if $x\in M$ and a basis $\{E_{i}\}$ in the tangent to $M$ bundle (i.e. if $\{E_{i}(z)\}$ is a basis in $T_{z}(M))$ are fixed, then the components $h^{i}(x,y)$ of $h(x,y)=:h^{i}(x,y)E_{i}(x)$ are local coordinates of every $y$, i.e. the map $y\to (h^{1}(x,y),\ldots  ,h^{n}(x,y)\in {\Bbb R}^{n}$is a local coordinate system on M. In this sense $h(x,y)$ may be called a vector coordinate of y.

 As a simple corollaries of (2.1), we find
\[
 h(x,y)=h(x,z)+L_{z  \to x}h(z,y),\qquad (2.5)
\]
\[
h(x,y)=-L_{y  \to x}h(y,x).\qquad (2.6)
\]

{\bf Remark.} If there is not a single coordinate
neighborhood containing $x$ and $y$, then the displacement vector depends
on the path $\gamma $. For instance, if this is the case and  there  exist
neighborhoods  $U^\prime \ni x, U^{\prime\prime}\ni y$ and
$U^\prime \cap U^{\prime\prime}\neq\emptyset$, then using in $U^\prime $ and
$U^{\prime\prime}$ coordinates like  those  in (2.4), we, writing explicitly
the dependence on $\gamma $, find
\[
h^{i^\prime }(x,y;\gamma )
=h^{i^\prime }(x,z)+\delta ^{i^\prime }_{j^\prime }
\frac{\partial z^{j'}}{\partial z^{j^{\prime\prime}} }
h^{j^{\prime\prime}}(z,y)
\]
\[
=z^{i^\prime }-x^{i^\prime }+\delta ^{i^\prime }_{j^\prime }
\frac{\partial z^{j'}}{\partial z^{j^{\prime\prime}} }
(y^{j^{\prime\prime}}-x^{j^{\prime\prime}}),
\]
 where $z\in \gamma (J)\cap U^\prime\cap U^{\prime\prime}$. From here is
evident the explicit dependence of the displacement vector on the path
entering in its definition (2.1) in the considered concrete case.

 Let $M$ be $a 4$-dimensional space-time. Let us consider a physical system
with a (contravariant) energy-momentum tensor $T^{ij}. ($The concrete
structure of $T^{ij}$or its dependence on other quantities, physical or
geometrical fields, is insignificant.) Let $\Sigma $ be a (time-like, if
there is a metric) hypersurface with a measure $d\Sigma _{k}=\epsilon
_{\hbox{kijl}}dx^{i}_{1}dx^{j}_{2}dx^{l}_{3}, \epsilon _{\hbox{kijl}}$being
the 4-dimensional antisymmetric $\epsilon $-symbols and $dx^{i}_{1},
dx^{j}_{2}$and $dx^{l}_{3}$being three linearly independent displacements on
$\Sigma $.

  We define the (4-)vector of energy-momentum of the system as
\[
p^{i}(x):=\frac{1}{c}
  \int_\Sigma H^{i}_{.j}(x,y)T^{jk}(y)d\Sigma _{k}(y)\qquad
(2.7)
\]
 in which $c=$const is the light velocity in vacuum.

 As a corollary of this definition (see also $eq. (2.5)$ from [12]), we get
\[
p(z)=L_{x  \to z}p(x).\qquad (2.8)
\]
 Let us define the tensor $P$ by
\[
 P^{ij}(x)
:=\frac{1}{c}\int_{\Sigma}
h^{i}(x,y)H^{j}_{.k}(x,y)T^{kl}(y)d\Sigma _{l}(y),\qquad
(2.9)
\]
 the antisymmetric part of which,
\[
 L^{ij}(x):=2P^{[ij]}(x):=P^{ij}(x)-P^{ji}(x),\qquad (2.10)
\]
is the orbital angular momentum tensor [2,3,10] of  the  investigated physical system. The fact that $L$ isn't a conserved  quantity  [3]  is not significant for the following. (A conserved quantity is the total angular momentum, which is a sum of $L$ and the spin  angular  momentum tensor [2,3,10].)

 Substituting (2.5) into (2.9), we get
\[
P(x)=h(x,z)\otimes p(x)+L_{z  \to x}P(z),\qquad (2.11)
\]
 which in a case of orbital angular momentum reduces to
\[
 L(x)=h(x,z)\wedge p(x)+L_{z  \to x}L(z),\qquad (2.12)
\]
where $\otimes $ is the  tensor  product  sign  and $\wedge$  is  the
antisymmetric external (wedge) product sign.

At the end of this section we shall write the expressions for the components of $h, p$ and $P$ in some special bases.

By propositions 4.1 and 4.2 from [12] (see also proposition 2.1 and the
assumption before it) there is a local holonomic basis (coordinate system) in
which the components of the bivector $H(x,y)$, representing $L$ in it, are
Kronecker's deltas, i.e. $H^{i}_{.j}(x,y)=\delta ^{i}_{j}$. In this basis,
from the definitions of $h, p$ and $P$, we find:
\[
 h^{i}(x,y)=y^{i}-x^{i},\qquad (2.13)
\]
\[
p^{i}(x)
=\frac{1}{c} \int_{\Sigma} T^{ik}(y)d\Sigma _{k}(y)=const,  \qquad (2.14)
\]
\[
P^{ij}(x)
=\frac{1}{c}\int_{\Sigma} (y^{i}-x^{i})T^{jk}(y)d\Sigma _{k}(y)
=P^{ij}(\mathbf{0} )-x^{i}p^{j}(x),\qquad (2.15)
\]
with $\mathbf{0}$ being the point with zero coordinates in the used basis.

Because of $p^{i}(x)=$const, from the used basis by linear transformation
with constant coefficients can be obtained a local holonomic basis with the
above-pointed property (see proposition 4.1 of [12]) in which
\[
 p^{i}=cM\delta ^{i}_{0},\qquad (2.16)
\]
 where
\[
 M:=\frac{1}{c}\int_{\Sigma} T^{0k}(y)d\Sigma _{k}(y)=const,  \qquad (2.17)
\]
 is the total mass of the investigated  physical  system.  Let's  note that
if the space is endowed with a metric and $x^{0}$is interpreted as a time
(coordinate), then (2.16) expresses the simple fact that $p(x)$ is a
time-like vector.

 So, in this basis
\[
 P^{ij}(x)=P^{ij}(\mathbf{0})-cMx^{i}\delta ^{j}_{0}.\qquad (2.18)
\]
 And, at  last, if we choose the hypersurface  $\Sigma $  as
$y^{0}=z^{0}=$const, then $d\Sigma _{k}(y)=\delta ^{0}_{k}d^{3}y$ and
$P^{0j}(\mathbf{0})=$cMz$^{0}\delta ^{j}_{0}$. Hence, we have
\[
  p^{0}(x)=cM=\frac{1}{c}
\int_{y^0=z^0} T^{00}(y)d^{3}y, p^{\alpha }(x)
= \frac{1}{c} \int_{y^0=z^0} T^{\alpha 0}(y)d^{3}y=0,
 \qquad (2.19)
\]
\[
 P^{0j}(x)=cM(z^{0}-x^{0})\delta ^{j}_{0},
 \qquad
 P^{\alpha j}(x)
=P^{\alpha j}(\mathbf{ 0})-cMx^{\alpha }\delta ^{j}_{0}.\qquad (2.20)
\]

\medskip
\medskip

 {\bf 3. CENTRE OF MASS IN A DISCRETE CASE}

\medskip
Let us have particles with masses $m_{a}$situated at some moment $t$ at the points $x_{a}$, where $a=1,\ldots  ,N$  numbers the particles. Below we suppose the particles total mass to be nonzero, i.e.   $m_{a}\neq 0$. Let $x$ be a fixed space-time point and the displacement vector $h(x,x_{a})$ be well defined (see the previous section).

{\bf Definition 3.1.} The mass centre of the masses $m_{a}$with respect to
the reference point $x$ at the moment $t$ is the point $x_{M}$ such that
\[
h(x,x_{M}):=
\Bigl( \sum_{a} m_{a} h(x,x_{a}) \bigr)
\Bigl( \sum_a m_{a  }\Bigr)^{-1}.\qquad (3.1)
\]

{\bf Remark.} With the change of time $t$ the point $x_{M}$describes a world
line, the world line of the system's mass centre.

As a consequence of (2.5), the mass centers $x_{M}$ and $y_{M}$ with respect
to the reference points $x$ and $y$ respectively are connected by
\[
 h(y,y_{M})=h(y,x)+L_{x  \to y}h(x,x_{M}).\qquad (3.2)
\]
In a local holonomic  basis in which $H^{i}_{.j}(x,y)=\delta ^{i}_{j}$, from
(3.1),  we easily get
\[
 x^{i}_{M}
= \Bigl(\sum_a m_{a}x^{i    }_{a    }\Bigr)
\Bigl( \sum_a m_{a  }\Bigr)^{-1}.\qquad (3.3)
\]

 {\bf Example 3.1} (Special  relativity;  $cf. [5])$.  Let  us  have  a
Minkowski's space-time $M^{4}$referred to Minkowskian coordinates.  As  a
concrete realization of the transport $L$ we shall  use  the  (pseudo-)
Euclidean transport defined by $H^{i}_{.j}(x,y)=\delta ^{i}_{j},
i,j=0,1,2,3$. The coordinates of any event $x\in M^{4}$are of the form
$(ct,{\bf x})$, where $c$ is the  velocity of light, $t$ is the time in the
used  frame  and ${\bf x}:=(x^{1},x^{2},x^{3})$, which may depend on $t$, is
the special coordinate of x.

So, in this case $(2.13)-(2.15)$ and (3.3) are valid. The last of  these
equality, due to $x^{0}_{a}=ct$ for every event, reduces to
\[
x^{0}_{M}=ct, \qquad
{\bf x}_{M} =
\Bigl(\sum_a m_{a}{\bf x}^{i    }_{a    }\Bigr)
\Bigl( \sum_a m_{a  }\Bigr)^{-1}
.\qquad (3.4)
\]
 If   we   define   $m_{a}$   as    $m_{a}={\cal E}_{a}/c^{2}$    where
${\cal E}_{a}=c^{2}m_{a}=c^{2}m^{0}_{a}     (1-(d{\bf x}/dt)^{2})^{-1/2},
m^{0}_{a}$ being the rest mass of the a-th  particle,   is the energy of the
a-th particle, we find
$x_{M}= \Bigl(\sum_a {\cal E}_{a}{\bf x}^{i    }_{a    }\Bigr)
\Bigl( \sum_a  {\cal E}_{a  }\Bigr)^{-1}$
Thus we can make the
inference that in the discrete case in special  relativity our definition
3.1 of mass centre reduces to the  known  classical one (see e.g. $[5],
ch.2, \S14)$. Let's note that the so  obtained mass centre depends on the
used basis (frame of reference), as ${\cal E}_{a}$are such quantities. If we
wish to get an  invariant  definition  of  $x_{M}$, then instead of
$m_{a}={\cal E}_{a}/c^{2}$we have to take $m_{a}=m^{0}_{a}$.

Now we want to show that in the discrete case there exists a very important for the following section connection between the mass centre $x_{M}$and the tensor $P$ with local components (2.9).

To begin with, let us remember that the component $T^{00}(z)$ of an
energy-momentum tensor is regarded as a energy density at $z [5]$. Hence it
can be written as $T^{00}(z)=c^{2}\rho (z), \rho (z)$ being the mass density
at $z$, which in the discrete case is
\[
 \rho (z)=\sum_a m_{a}\delta ^{3}({\bf x}_{a}-{\bf z}_{a}),\qquad (3.5)
\]
 where $\delta ^{3}$is the 3-dimensional Dirac's delta function.

If a local holonomic basis in which $H^{i}_{.j}(x,y)=\delta ^{i}_{j}$is used
and $\Sigma $ is defined by $y^{0}=z^{0}=$const, then $d\Sigma _{k}(y)=\delta
^{0}_{k}d^{3}y$ and from (2.9), we obtain
\[
 P^{i0}(x)
= \frac{1}{c}\int_{y^0=z^0} h^{i}(x,y)T^{00}(y)d^{3}y
=  \frac{1}{c}\int_{y^0=z^0}  h^{i}(x,y)\sum m_{a}\delta ^{3}({\bf
x}_{a}-{\bf y})d^{3}y
\]
\[
 =c\sum_a m_{a}h^{i}(x,x_{a})|_{x^{0}_a=z^0}
\]
which may also be written as
\[
P^{00}(x)=cM(z^{0}-x^{0}),
  \qquad
P^{\alpha 0}(x)=cMh^{\alpha }(x,x_{M})|_{x^{0}_M=z^0},\qquad (3.6)
\]
where $(cf. (2.17))$ the total mass of the system is
\[
M:= \int_{y^0=z^0}  \rho (y)d^{3}y
=\sum_a m_{a}.\qquad (3.7)
\]
 Analogous calculations (see (2.14) and (2.15)) show that:
\[
  p^{0}(x)
= \frac{1}{c}\int_{y^0=z^0}  T^{00}(y)d^{3}y
=cM,
 \qquad
p^{\alpha }(x)= \frac{1}{c}\int_{y^0=z^0} T^{\alpha 0}(y)d^{3}y,\qquad (3.8)
\]
\[
P^{0\alpha }(x)=(z^{0}-x^{0})p^{\alpha }(x),
 \qquad
P^{\alpha \beta }(x)= \frac{1}{c}\int_{y^0=z^0}
y^{\alpha }T^{\beta 0}(y)d^{3}y-cx^{\alpha }p^{\beta }(x). (3.9)
\]
The  most important for the following result here is the  connection
between $h(x,x_{M})$ and $P$ expressed explicitly by (3.6).

\medskip
\medskip
\medskip
 {\bf 4. CENTRE OF MASS: GENERAL CASE}
\nopagebreak

\medskip
The conclusion from the previous section is that the  mass centre of a physical system (if it exists!) must be connected with the tensor $P$ and in the discrete case this connection must reduce to the already established one.

So, we state the problem for expressing in a covariant way $h(x,x_{M})$
through $P(x_{M})$. Due to (3.15) these quantities are connected by the
relation
\[
P^{ij}(x_{M})
=H^{i}_{.k}(x_{M},x)H^{j}_{.l}(x_{M},x)
P^{kl}(x)+h^{i}(x_{M},x)p^{j}(x_{M}),\qquad (4.1)
\]
 which is a simple corollary from the  corresponding  definitions  and directly can't serve as
an equation for determination of
\[
 h(x_{M},x)=-L_{x  \to x_{M}}h(x,x_{M}).\qquad (4.2)
\]
 Hence, to express  $h(x_{M},x)$ through $P(x_{M})$ we must impose on the
latter a certain number of independent conditions such that by the usage  of
(4.1) they must be solvable with respect to (some of) the  components of
$h(x_{M},x)$ and such that the so obtained  dependence  in  a  discrete case
must coincide with the one established in section 3. The type of these
conditions is sufficiently arbitrary and this is the cause  for the possible
existence of different inequivalent definitions of  the mass centre on the
basis of $P$ or  the  orbital  and/or  spin  angular momentum, all of which
in the corresponding special cases  reduce  to its classical definition.
Below we analyze only the linear conditions that can be imposed on $P$ which
most of all fit to the general  spirit of tensor calculus and general
relativity.

The general form of the mentioned linear conditions is
$B^{i}_{jk}(x_{M})P^{jk}(x_{M})=b^{i}(x_{M})$ for some tensors $B^{i}_{jk}$
and $b^{i}$, i.e.
\[
B^{i}_{jk}(x_{M})H^{j}_{.l}(x_{M},x)H^{k}_{.n}(x_{M},x)P^{\ln}(x) +
B^{i}_{jk}(x_{M})h^{j}(x_{M},x)p^{k}(x_{M})=b^{i}(x_{M}).
 \quad (4.3)
\]
In section 3
we saw that only $h^{\alpha }, \alpha =1,\ldots  ,n-1=3$ are connected with
$P$, the component $h^{0}$being independent of it. Hence only $n-1=3$ of
these $n=4$ conditions must be independent, i.e.
\[
\det[B^{i}_{jk}(x)p^{k}(x)]=0,
 \qquad  i,j,k=0,1,\ldots  ,n-1=3,\qquad (4.4)
\]
\[
\det[B^{\alpha }_{\beta k}(x)p^{k}(x)] \neq 0,
 \qquad  \alpha ,\beta ,\gamma=1,\ldots  ,n-1=3.\qquad (4.5)
\]
(The last condition may  always  be fulfilled  with  an  appropriate
renumbering of $B^{i}_{jk}(x).)$

The condition (4.4) is equivalent to  the existence of nonvanishing covector
field $q$ such that
\[
 B^{i}_{jk}(x)q_{i}(x)p^{k}(x)=0
 \qquad
{\bigl(}\sum(q_{i}(x))^{2}\neq 0{\bigr)}.\qquad (4.6)
\]
 On the opposite, if we fix  a  covector  field  $q\neq 0$
and  define $B^{i}_{jk}(x)$ as any solution of $(4.5)-(4.6)$, we shall obtain
some  relation (4.3) satisfying the needed conditions.

 Let there be given a nonvanishing covector field q. It is easily verified
that the quantities $^{0}B^{i}_{jk}(x):=2\delta ^{[i}_{j}\delta
^{l]}_{k}q_{l}(x)=(\delta ^{i}_{j}\delta ^{l}_{k}-\delta ^{l}_{j}\delta
^{i}_{k})q_{l}(x)$ satisfy all of the above conditions. So, putting
$B^{i}_{jk}(x)=:^{0}B^{i}_{jk}(x)+^\prime B^{i}_{jk}(x)$ and
$Q^{i}(x):=-^\prime B^{i}_{jk}(x)P^{jk}(x)+b^{i}(x)$ into (4.3), we see that
$h(x_{M},x)$ must be a solution of
$P^{[jk]}(x_{M})q_{k}(x_{M})=Q^{j}(x_{M})$, or
\[
q_{k}(x_{M})H^{j}_{.l}(x_{M},x)H^{k}_{.n}(x_{M},x)
P^{[\ln]}(x)+q_{k}(x_{M})h^{[j}(x_{M},x)p^{k]}(x_{M})
= Q^{j}(x_{M}) ,
\]
where $q$, $Q$ and $p$ must satisfy the conditions
\[
 Q^{i}(x)q_{i}(x)=0,
 \qquad  p^{i}(x)q_{i}(x)\neq 0.\qquad (4.8)
\]
 The former of them
is a corollary  from  (4.6)  and  the  latter  one ensures the solvability of
(4.7) with respect to  $h(x_{M},x)$  in  space- times with dimension greater
then one. (Evidently,  if  $p^{i}(x)q_{i}(x)=0$, from (4.7) can be obtained
no  more  then   the  linear  combination $q_{i}(x_{M})h^{i}(x_{M},x)$, but
not $h^{i}(x_{M},x)$ itself.)

 From (4.7), we get
\[
h^{i}(x_{M},x)
= \frac{1}{q_k(x_M) p^k(x_M)}
\bigl[ Q^{i}(x_{M}) + (q_{k}(x_{M})h^{k}(x_{M},x))p^{i}(x_{M})
\]
\[
 - 2H^{i}_{.l}(x_{M},x)H^{k}_{.n}(x_{M},x)P^{[\ln]}(x)q_{k}(x_{M})
\bigr] .\qquad (4.9)
\]
 Let us investigate this expression.

Firstly, (4.9) defines only the spacial components $h^{\alpha }(x_{M},x)$,
leaving the time component $h^{0}(x_{M},x)$ arbitrary. To prove this, let's
take a basis $\{E_{i}\}$ such that $H^{i}_{.j}(x,y)=\delta ^{i}_{j}$. In it
(4.9) reduces to
\[
h^{\alpha }(x_{M},x)
= \frac{1}{q_k(x_M) p^k(x_M)}
  \bigl[Q^{\alpha }(x_{M}) + (q_{k}(x_{M})h^{k}(x_{M},x))p^{\alpha
}(x_{M})
\]
\[
- 2P^{[\alpha k]}(x)q_{k}(x_{M})\bigr],\qquad (4.10a)
\]
\[
h^{0}(x_{M},x)
= \frac{1}{q_k(x_M) p^k(x_M)}
\bigl[Q^{0}(x_{M})
+ (q_{k}(x_{M})h^{k}(x_{M},x))p^{0}(x_{M})
\]
\[
-  2P^{[0\beta ]}(x)q_{\beta }(x_{M})\bigr].\qquad (4.10b)
\]
 As $q\neq 0$, for some i we  must have $q_{i}(x_{M})\neq 0$. Let, e.g.,
$q_{0}(x_{M})\neq 0$.

Substituting (4.10a) and $Q^{\alpha }(x_{M})q_{\alpha }(x_{M})=-Q^{0}(x_{M})q_{0}(x_{M}) ($see (4.8)) into $h^{0}(x_{M},x)\equiv (q_{0}(x_{M}))^{-1}[q_{k}(x_{M})h^{k}(x_{M},x)-h^{\alpha }(x_{M})q_{\alpha }(x_{M})]$, we obtain (4.10b). So, (4.10b) is a consequence of (4.10a). Evidently, the same result is true if $q_{i}(x_{M})\neq 0$ for some other fixed value of i.

Now we shall study what conditions must satisfy $q$ and $Q$ if in the discrete case the right hand side of (4.10a) reproduces the same result as (3.1).

 For simplicity and brevity a basis $\{E_{i}\}$ in which
\[
 H^{i}_{.j}(x,y)=\delta ^{i}_{j},
\quad
 p^{i}(x)=\frac{1}{C} \int_{y^0=z^0}
	T^{i0}(y)d^{3}y=cM\delta ^{i}_{0},
 \quad
M=const \neq 0
 \quad (4.11)
\]
 will be used. In it (4.10a) gives
\[
^{\alpha }(x_{M},x)
= \frac{1}{cM q_0(x_M)}
[Q^{\alpha }(x_{M})-2P^{[\alpha 0]}(x)q_{0}(x_{M})-2P^{[\alpha \beta
]}(x)q_{\beta }(x_{M})].
 \qquad (4.12)
\]
(In this basis $q_{0}(x_{M})\neq 0$ because  of (4.8) and (4.11).)

Substituting, in accordance with (4.11) and $(3.6)-(3.9)$,
here
$P^{\alpha 0}(x)
=cMh^{\alpha }_{0}(x,x_{M})
=-cMh^{\alpha }_{0}(x_{M},x), P^{0\alpha }(x)=0$
and $P^{\alpha \beta }(x)=P^{\alpha \beta }(\mathbf{ 0})$, where $h_{0}$ is
defined by the right hand side of (3.1), we find
\[
 h^{\alpha }(x_{M},x)
=h^{\alpha }_{0}(x_{M},x)
+  \frac{1}{cM q_0(x_M)}
[Q^{\alpha }(x_{M})-2P^{[\alpha \beta ]}(\mathbf{ 0})q_{\beta }(x_{M})].
 \qquad (4.13)
\]
Therefore
\[
h^{\alpha }(x_{M},x)=h^{\alpha }_{0}(x_{M},x),\qquad (4.14)
\]
 as we must
have, if and only if $Q^{\alpha }(x_{M})=2P^{[\alpha \beta ]}(\mathbf{
0})q_{\beta }(x_{M})$, from  which, due to (4.8) and $q_{0}(x)\neq 0$,
follows $Q^{0}(x)=0$, i.e. (4.14) is  equivalent to
\[
 Q^{i}(x_{M})=2\delta ^{i}_{\alpha }H^{\alpha }_{.j}(x_{M},\mathbf{ 0})H^{\beta
}_{.l}(x_{M},\mathbf{ 0})P^{[jl]}(\mathbf{ 0})q_{\beta }(x_{M}).\qquad (4.15)
\]
 Hence
$Q$ must depend linearly upon $q$ and the  antisymmetric  part of P. But,
because the Greek indices don't take the  value  zero,  it depends also on
the used basis $\{E_{i}\}$ which isn't uniquely  defined  by the conditions
$H^{i}_{.j}(x,y)=\delta ^{i}_{j}$and $p^{i}(x)=cM\delta ^{i}_{0}. ($These
conditions fix $\{E_{i}\}$ up  to  a  transformation  with  a  constant
nondegenerate  diagonal matrix.) The only way to be skipped that last
dependence is to  admit that in $\{E_{i}\}$ is fulfilled $q_{\alpha
}(x_{M})=0$, or
\[
 q_{i}(x_{M})=q_{0}(x_{M})\delta ^{0}_{i}, q_{0}(x_{M})\neq 0,\qquad (4.16)
\]
which implies (see (4.15))
\[
 Q^{i}(x_{M})=0.\qquad (4.17)
\]
 The above discussion and its results can be summarized into

{\bf Proposition 4.1.} Let $h(x_{M},x)$ depends linearly on $P(x_{M})$ and
in the discrete case reduces to (3.1). Let there be chosen a nonvanishing
covector field q. Let there exists a local holonomic basis such that in it:
\[
 H^{i}_{.j}(x,y)=\delta ^{i}_{j},\qquad (4.18a)
\]
\[
p^{i}(x)=cM\delta^{i}_{0},  \qquad M=const\neq 0,\qquad (4.18b)
\]
\[
q_{i}(x)=q_{0}(x)\delta ^{0}_{i}, \qquad  q_{0}(x)\neq 0.\qquad (4.18c)
\]
 Then in any basis the spacial
coordinates  $x^{\alpha }_{M}$  of  $x_{M}$  are  uniquely defined by the
equation
\[
P^{[ik]}(x_{M})q_{k}(x_{M})=0,\qquad (4.19)
\]
 or, equivalently, by
\[
 h^{i}(x_{M},x)
=\frac{1}{(q(p))(x_M)}
  [(q(x_{M})(h(x_{M},x)))p^{i}(x_{M})
\]
\[
-2 H^{i}_{.k}(x_{M},x)H^{j}_{.l}(x_{M},x)P^{[kl]}(x)q_{j}(x_{M})],\qquad
(4.20)
\]
 which leaves the time component $x^{0}$of $x_{M}$ in the above
special  basis arbitrary.

Let us turn now our attention on the covector field $q$, which must satisfy only the condition (4.18c). In this connection are important the following two observations. Firstly, the defined by (4.19) mass centre $x_{M}$, generally, depends on the choice of $q$ which is in a great extend arbitrary and until now hasn't any physical meaning. Secondly, the equations (4.19), as well as the results leading to
proposition 4.1, imply the existence of some dependence of $q$ on p. These two facts, the above-considered discrete case and the investigations in [2,3] are a hint for us to propose the following general definition of mass centre.

 Let the space-time be endowed with a linear transport $L$ and independently
with a metric $g$ with covariant components $g_{ij}=g_{ji}$ and signature
$(+---)$. Then, roughly speaking, the mass centre $x_{M}$is defined by
proposition 4.1 with $q_{i}=g_{ij}p^{j}$. More precisely, we give

{\bf Definition 4.1.} The mass centre of a system described by an energy-momentum tensor is the unique point $x_{M}$satisfying the following three conditions:

 1. At the point $x_{M}$in any local basis is valid the equation
\[
 P^{[ik]}(x_{M})g_{kl}(x_{M})p^{l}(x_{M})=0.\qquad (4.21)
\]

 2. In a
neighborhood of $x_{M}$there exist  local  coordinates  $\{x^{i}\}$ such that
in the associated to them basis $\{\partial /\partial x^{i}\}$ to be
fulfilled:
\[
 H^{i}_{.j}(x_{M},y)=\delta ^{i}_{j},\qquad (4.22a)
\]
\[
p^{i}(x_{M})=cM\delta ^{i}_{0}, \qquad  M=const\neq 0,\qquad (4.22b)
\]
\[
g_{i0}(x_{M})=g_{00}(x_{M})\delta ^{0}_{i},
 \qquad  g_{00}(x_{M})\neq 0.\qquad(4.22c)
\]

3. In the coordinates in which (4.22) hold the time component of $x_{M}$is
$x^{o}_{M}=ct, t$ being the time in these coordinates.

\medskip
 {\bf 5. Comments}

\medskip
 Now we shall make some remarks concerning definition 4.1.

Firstly, the equation (4.21) is a special case of $eq. (4.19)$ when the choice $q_{i}=g_{ij}p^{j}$is made. Our opinion is that this connection between $q$ and $p ($in a metric space-time) is the only "reasonable" one which prevents the dependence of $x_{M}$on a sufficiently arbitrary quantity q. Moreover, in this way is given a physical meaning to $q$ as the covector (1-form) corresponding by means of the metric to the momentum p. This is important because by its meaning the mass centre must depend only on t

Secondly, the conditions (4.22) ensure the solvability of $eq. (4.21)$ with respect to $x_{M}$and the coincidence in the discrete case of the so obtained value of $x_{M}$with the one obtained independently by definition 3.1. Let's also note that the condition (4.22c) is a
simple corollary of $q_{i}=g_{ij}p^{j}$and (4.18c).

Thirdly, the three conditions (4.22) have a different meaning and in the
general case of arbitrary metric they can't be satisfied simultaneously. The
first of them, (4.22a), expresses the fact that the associated to the used
transport connection is torsion free (in addition to its zero curvature). The
second one, (4.22b), shows that the (linear) momentum $p$ is (by definition)
a time-like vector and that its direction is taken as a direction of the time
(zeroth) coordinate axes, which is possible because of $M\neq 0. ($If $M=0$,
then $x$ is left completely arbitrary by (4.21) and (4.22), i.e. for massless
systems any space-time point can serve as their mass centre.) These two
conditions are always compatible in accordance with propositions $4.1-4.3$ of
[12] as in a basis in which (4.22a) is valid it is fulfilled (2.14). The last
condition, (4.22c), enables us to interpret $x^{0}_{M}/c=t$ as a time in the
described frame of reference (if it exists). This condition is very
restrictive one. In fact, if $x_{M}$was a fixed point, then with a linear
transformation with constant coefficients it is possible (see [12],
proposition 4.1) to transform the basis in which (4.22a) holds into a basis
in which (4.22a) and (4.22c) are valid simultaneously. But, generally, in
such a basis (4.22b) will not be satisfied. Moreover, as $x_{M}$describes
with the change of time a whole world line, the mass centre's world line, in
the general case one needs a linear transformation with nonconstant
coefficients to satisfy (4.22c) and if this is the real situation, then, by
[12], proposition 4.1, in the new basis the property (4.22a) will be lost.
The conclusion from these considerations is that (4.22c) puts a significant
restriction on the possible metrics which are admitalbe if we want to be well
defined the mass centre (world line) of an arbitrary material system. In
short, in a given space-time the equation (4.21) defines a mass centre (world
line(s)) if and only if all of the conditions (4.22) can be satisfied in some
local holonomic basis.

Fourthly, as it was proved above, the equation (4.21) and the conditions
(4.22) define, in a basis in which (4.22) are satisfied, only the spacial
coordinates $x^{\alpha }_{M}$of the mass centre $x_{M}$, but its time
coordinate $x^{0}_{M}$is left by them completely arbitrary. This last
component is fixed by the third condition of definition 4.1 in such a way as
to give its appropriate value in the discrete and classical cases.

Fifthly, by (4.12) with $Q=0$ and $q_{i}=g_{ij}p^{j}$, in the basis
described by (4.22) the mass centre has the following coordinates
\[
 x^{0}_{M}=ct,\quad
x^{\alpha }_{M} = x +\frac{1}{cM} P^{[\alpha 0]}(x).\qquad (5.1)
\]
 In any other basis the coordinates of $x_{M}$can  be  obtained  from the
components of the displacement vector $h(x,x_{M})$ in this basis.

Sixthly, in $[2-4]$ to define the mass centre a "similar" to (4.21) equation is proposed in which the orbital angular momentum $L^{ij}:=P^{[ij]}$is replace with the total angular momentum $J^{ij}=L^{ij}+S^{ij}$which includes the spin angular momentum $S^{ij}. ($In [3,4] the bitensor $H(x,y)$ is replaced with another bitensor and the case of general relativity is considered, but this circumstances are insignificant now.) The only reason for this being that $J^{ij}$is a conserved quantity. We consider this def

Seventhly, in $[5], ch$. II, \S14 is pointed that the presented therein definition of mass centre gives different points for it in different frames (bases), i.e. it depends explicitly on the used local coordinates, even in the simple case of special relativity (cf. our example 3.1). Evidently, our definition 4.1 is free of this deficiency the cause for this being the condition (4.22c) (see also (4.18c) and (4.17)) and the general usage of the displacement vector $h(x,x_{M})$ for the definition of $x_{M}($see

At the end, the above discussion can be summarize as follows. If in a space-times endowed with a (flat) linear transport (connection) and a metric we admit a linear relationship between $P(x_{M})$ and $h(x,x_{M})$, then the mass centre (mass centre's world line) is well defined by definition 4.1 and it exists if the conditions (4.22) can be satisfied in some local coordinates. It is important to be noted that just this is the classical case of special relativity.

\medskip
 {\bf References}

\medskip
1. Bailey I., W. Israel, Relativistic dynamics of extended bodies and polarized media: An eccentric approach, Annals of Physics. vol.
${\bf 1}{\bf 3}{\bf 0}, pp. 188-214, 1980$.\par
2. Dixon W.G., A covariant multipole formalism for extended test bodies in general relativity, Nuovo cimento, vol. {\bf 34}, No. $2, pp. 317-339, 1964$.\par
3. Dixon W. G., Dynamics of extended bodies in general relativity. I. Momentum and angular momentum, Proc. Roy. Soc. London, vol. ${\bf A} {\bf 3}{\bf 1}{\bf 4}{\bf ,} pp. 499-527., 1970$.\par
4. Dixon W.G., Extended bodies in general relativity: Their description and motion, in: Isolated gravitating systems in general relativity, North-Holland Publ. Co., Amsterdam, 1979.\par
5. Landau L. D., E. M. Lifshiz, The Classical Theory of Fields, Addison-Wesley, Reading, Mass., 1959.\par
6. Lovelock D., H. Rund, Tensors, Differential Forms, and  Variational Principals, Wiley-Interscience Publication, John Wiley \& Sons, New York-London-Sydney-Toronto, 1975.\par
7. Kobayashi S., K. Nomizu, Foundations of differential geometry, vol. 1, Interscience Publishers, New York-London, 1963.\par
8. Ruse H. S., Proc. London Math. Soc., vol. {\bf 32}, p. $87, 1931 ($Quart. J. Math., vol. 2, p. $190, 1931)$.\par
 9. Synge J. L., Proc. London Math. Soc., vol. {\bf 32}, p. $241, 1931$. \par
10. Synge J. L., Relativity: The General Theory, North-Holland Publ. Co., Amsterdam, 1960.\par
11. de Witt B. S., R. Brehme, Radiation dumping in a gravitational fields, Annals of Physics, vol. ${\bf 9}, pp. 220-259, 1960$.\par
12. Iliev B. Z., Flat linear connections in terms of flat linear transports in tensor bundles, Communication JINR, $E5-92-544$, Dubna, 1992.\par
13. Iliev B.Z., Linear transports along paths in vector bundles. I. General theory, Communication JINR, $E5-93-239$, Dubna, 1993; Transports along paths in fibre bundles. General theory, Communication JINR, $E5-93-299$, Dubna, 1993.

\newpage

\medskip
\medskip
\noindent Iliev B. Z.\\[5ex]

\medskip
 \noindent
Centre of Mass in Spaces with Torsion Free Flat Linear Connection\\[5ex]

\medskip
\medskip
\medskip
The concept "centre of mass" is analyzed in spaces with torsion free flat
linear connection. It is shown that under sufficiently general conditions it
is almost uniquely defined, the corresponding arbitrariness in its definition
being explicitly described.\\[5ex]

\medskip
\medskip
\medskip
\medskip
The investigation has been  performed  at  the  Laboratory  of Computing
Techniques and Automation, JINR.

\end{document}